\documentstyle[preprint,pre,aps]{revtex}
\begin{document}
\draft

\title{Time dependent Ginzburg-Landau equation for
an N-component model of self-assembled fluids}

\author{Umberto Marini Bettolo Marconi}
\address{Dipartimento di Matematica e Fisica, Universit\`a di Camerino,
         Via Madonna delle Carceri,I-62032 , Camerino, Italy}

\author{Federico Corberi}
\address{Dipartimento di Fisica Teorica, Universit\`a ``Federico II'',
         Mostra d'Oltremare, Padiglione 19, I-80125, Napoli, Italy}
\maketitle
\begin{abstract}
 We study the time evolution of an N-component model of bicontinuous
microemulsions
based on a time dependent Ginzburg-Landau equation quenched
from an high temperature uncorrelated state to the low
temperature phases.
 The behavior of the dynamical structure factor $\tilde C(k,t)$
is obtained, in each phase, in the framework of the large-$N$
limit with both conserved (COP) and non conserved
(NCOP) order parameter dynamics.
At zero temperature the system shows multiscaling in the unstructured region
up to the tricritical point for the COP whereas ordinary scaling
is obeyed for NCOP. In the structured phase, instead,
the conservation law is found to be irrelevant and the form
$\tilde C(k,t) \sim t^{\alpha / z} f((|k-k_m| t^{1/z})$,
with $\alpha=1$ and $z=2$, is obtained in every case.
 Simple scaling relations are also derived for the structure
factor as a function of the final temperature of the
thermal bath.

\end{abstract}


 Self-assembled fluids are mixtures of water, oil and surfactants
displaying a richer thermodynamic behaviour than ordinary fluids
because they appear to  have a tendency to order and
form patterns. In fact, if the surface tension can be made sufficiently
low, by increasing the surfactant concentration, one observes a microemulsion
phase stabilized by spontaneous
curvature and entropy of mixing, whereas at higher values of the surface
tension one obtains oil-rich  or water-rich regions.
By increasing further the surfactant concentration, however,
the fluid acquires an internal order as revealed by the several structured
phases.

 So far most of the experimental and theoretical studies have been
devoted to the investigation of the equilibrium properties of such systems
whereas the properties far from equilibrium of complex
fluids are not so well
understood \cite{Daw}.
 In the present Letter, by generalizing to a large number of components
the GL model for oil-water-surfactant
mixtures  put forward by Gompper-Schick \cite{GS},\cite{GHS},
we study the dynamics after a quench.

By associating an order parameter with the difference
in the density of water and oil molecules we build the GL functional
as the sum of a local term plus a non local interaction, which
is mathematically convenient to represent as a gradient expansion.
 The model is described by the following GL hamiltonian:
\begin{equation}
\label{eq:model}
   H[\bbox{\phi}(\bbox{ x})] = \int d{\bbox{x}}
[\frac{1}{2}(\nabla^2 \bbox{\phi})^2
+\frac{b}{2}(\nabla \bbox{\phi})^2
+\frac{w}{2N}( \bbox{\phi})^2 (\nabla \bbox{\phi})^2
+\frac{r}{2}( \bbox{\phi})^2
+\frac{g}{4N}(( \bbox{\phi})^2)^2]
\end{equation}
where $r$ and $g$ (with $g>0$) are the quadratic
and quartic couplings of the GL
theory.
The term containing the squared Laplacian  represents a curvature energy
contribution,
whereas the second and the third term are proportional to the surface
energy.
The
peculiarity of the self-assembled fluids is reflected by the presence
of higher order derivatives in the gradient expansion and by
the sign of
the coefficient $b$ of the  square gradient term, which can be either
positive as in ordinary fluids or negative according to
the value of the thermodynamic  control parameters, as found from the
fits of the experimental data \cite{Teub},
while the constant $w$ is assumed to be positive.

Incidentally, our results turn out to be of interest for a classical
problem of relaxation dynamics, apparently quite different as the hydrodynamic
fluctuation at the convective instability, the so called Swift-Hohenberg
model \cite{Bra}- \cite{Gold}.

Within the large $N$ limit
the equation of motion for the order parameter and its fluctuations
can be written in a closed form and we can extract explicitly the
exact asymptotic behaviour \cite{CZ}- \cite{Bray}.
 A continuum description of the model begins with associating the modulus of
the vector order parameter
$\bbox{ \phi}({\bbox x},t)=(\phi_1({\bbox x},t),..,\phi_N({\bbox x},t))$
with the
difference in the density of water and oil molecules and assuming that its
evolution towards equilibrium is described by the Langevin equation:
\begin{equation}
\label{eq:Lang}
   \frac{\partial \phi_{\alpha}(\bbox{ x},t)}{\partial t} =
               - \Gamma(\vec x)\frac{\delta}{\delta
\phi_{\alpha}(\bbox{ x},t)}\, H[\bbox{\phi}(\bbox{ x})]
               + \eta_{\alpha}(\bbox{x},t)
\end{equation}
Here $\vec \eta(\vec x,t)$ represents a Gaussian white noise
with zero average and
$<\eta_{\alpha}(\bbox{x},t) \eta_{\beta}(\bbox{x'},t)>= 2 T_f \Gamma(\bbox{x})
\delta_{\alpha,\beta} \delta (\bbox{x-x'}) \delta(t-t')$, where
$T_f$ is the temperature of the
final equilibrium
state and the kinetic coefficient takes a constant value $\Gamma$ for
non conserved order parameter (NCOP),
whereas it is given by $- \Gamma \nabla^2$ in the conserved case (COP).

 In order to study the properties of the system one considers the
long-range behaviour of the
equal time real space connected correlation
function $C(r,t)=<\phi_{\alpha}(R+r,t) \phi_{\alpha}(R,t)>$ and its
Fourier transform, the structure factor, $\tilde C(k,t)$, which are both
independent of the index $\alpha$ due to the internal symmetry.

 In the large $N$-limit from eq. (\ref{eq:Lang})
one derives the following evolution for the structure function :
\begin{equation}
\label{eq:ct}
\frac{d}{dt}\tilde{C}(k,t)= -2 \Gamma k^p
[ k^4  +\dot B(t) k^2+\dot Q(t) ] \tilde{C}(k,t)+2 k^p \Gamma T_f
\end{equation}
with $p=0$ for NCOP dynamics and $p=2$ for COP.
We have introduced the two auxiliary functions
$B(t)$ and $Q(t)$, which have to be determined self-consistently through:
\begin{equation}
\label{eq:bt}
\dot B(t)= b+w S_0(t)
\end{equation}
and
\begin{equation}
\label{eq:qt}
\dot Q(t)= r+g S_0(t)+ w S_2(t)
\end{equation}
where the integrals
\begin{equation}
\label{eq:S0}
S_0(t) =\int_{|k|<\Lambda} \frac{d^d k}{(2 \pi)^d} \tilde C(k,t)
\end{equation}
and
\begin{equation}
\label{eq:S2}
S_2(t) =\int_{|k|<\Lambda} \frac{d^d k}{(2 \pi)^d} k^2 \tilde C(k,t)
\end{equation}
contain a phenomenological momentum cutoff $\Lambda$.

 At equilibrium ( see also \cite{Gon}), when $r\le 0$, the system displays
a low temperature ordered, ' magnetic'
phase  with non vanishing
order parameter $<\bbox {\phi}>$ and a high temperature disordered
(paramagnetic) phase as shown in Fig. \ref{fig:fig1}.
In order to draw the phase diagram of the system we consider the
equilibrium value of the structure factor $\tilde C_{eq}(k)$ which reads:
\begin{equation}
\label{eq:Teub}
\tilde C_{eq}(k)=\frac{T_f}{k^4+b_r k^2+D},
\end{equation}
where we have introduced $b_r=\lim_{t \to \infty}(b+w S_0(t))$
and $D=\lim_{t \to \infty}(r+g S_0(t)+ w S_2(t))$.
 The 'ordered phase' $<\bbox{ \phi} \ne 0 >$ is bounded from above by the line
of 'magnetic'
critical points $T_c(b)$ shown in Fig.
\ref{fig:fig1} . It terminates at the
tricritical point ($b_L=r w /g$, $T_{tr}$), located at $T_{tr}=0$
in $d < 4$.
Points below such a curve correspond to a vanishing value of the parameter
$D$. In other words the line $T_c(b)$ separates a high temperature phase
with finite correlation length and finite fluctuations from a low temperature
magnetic phase with infinite correlation length and divergent fluctuations at
zero wavevector, due to the presence of massless Goldstone modes.

Within the disordered phase
the two-particle correlation function displays
two different behaviors according to the sign
of the discriminant $\Delta=[D-\frac{b_r^2}{4}]$.
In fact,
the vanishing of $\Delta$ identifies the so called disorder line, i.e.
the
borderline between a regime
(since it is not a proper thermodynamic phase) with monotonically
decaying correlations, which is the analogue of a paramagnetic phase,
and a phase with oscillatory-decaying
correlations, which in the present model corresponds to a
microemulsion phase.
The locus where
the structure factor, $\tilde{C}_{eq}(k)$ starts developing
a peak at finite wavevector $k$, when
$b_r(T_f)=0$, is named Lifshitz line. It is contained in the
microemulsion phase and terminates at the tricritical point.
To summarize, for positive values  of
$\Delta$,
$C(r) \sim \frac{e^{-r/ \xi}}{r} sin( k_m r)$
with $\frac{1}{\xi}=[\frac{1}{2}\sqrt{D}+\frac{b_r}{4}]^{1/2}$
and
with $k_m=[\frac{1}{2}\sqrt{D}-\frac{b_r}{4}]^{1/2}$. For
negative values of $\Delta$,  the correlation decays monotonically,
showing an Ornstein-Zernike behavior.
 Finally along the semiaxes $T_f=0$ and $b \leq r w/g $, $\Delta=0$
and the structure factor diverges at a finite wavevector.
Such a line is the remnant of the lamellar phase, which is
unstable for  any finite temperature, because the system does not
support
topological defects in any dimension $d$ less than $N$.

 We shall consider now the phase ordering of the present system
following a quench from an uncorrelated high temperature state
for different choices of $b_r$.
In order to characterize the quench process one needs to specify the
the set of parameters $b,w,r,g$ and
initial conditions for the field; this can be done by
assigning the structure factor at $t=0$. One usually assumes
the initial state to be uncorrelated and chooses the initial
condition $C(k,0) = \Delta_0=const$.
In the following we shall consider $r<0$.
 The asymptotic form of $\tilde C(k,t)$,
for the $T_f=0$ case, can be easily calculated by
imposing the matching of
$S_0(t)$ and $S_2(t)$, calculated by means of the general
solution of eq. (\ref{eq:ct})
with their equilibrium constant values.

 In the NCOP case and for positive values of $b_r$, since
the curvature term is
asymptotically irrelevant,
the structure factor has the usual scaling form as for
simple fluids
$\tilde C(k,t)\sim t^{\alpha/z}
f(k t^{1/z})$, with $f(x)=e^{-x^2}$, $\alpha=d$ and $z=2$,
for a quench below the critical temperature ($T_f< T_c$).
The exponent $z=2$ is, in fact, related to the growth of the
domain size $L(t)=(2 \Gamma t)^{1/2}$, which is controlled by
surface tension.
In the COP case one observes, instead, the multiscaling behaviour of
the structure factor \cite{CZ}, due to the existence of two
marginally different scaling lengths, $L(t)=(2 \Gamma t)^{1/4}$
and $k^{-1}_m(t) \sim (t/ \log t)^{1/4}$, where $k_m(t)$ is the position
of the peak of the structure factor $\tilde C (k,t)$. This results in
a multiscaling form for $\tilde C(k,t) \sim (k_m^{2-d}L^2)^{\phi(k/k_m)}$
with $\phi(x)=1-(1-x^2)^2$.

 Of particular interest is the study of a quench at the tricritical
point, where the 'magnetic', the 'paramagnetic' and the 'lamellar'
phase meet. This occurs
when $b_r$ vanishes, at $T_f=0$, i.e. in correspondence
of the bare value $b_L=rw/g$.

We study, first, the growth  of the structure factor upon approaching
$T_f=0$ along a particular Lifshitz line, by
considering the simpler case $w=0$, which yields $b_L=0$.
For the NCOP at $T_f=0$ the structure factor reads
$C(k,t) \sim \Delta_0 e^{-2\Gamma k^4 t} t^{d/4}$,
i.e. a scaling
form characterized by $\alpha=d$, $z=4$ and scaling
function $f(x)=e^{-x^4}$. The typical domain size evolves according to
$L(t)=(2 \Gamma t)^{1/4}$, being controlled by the curvature
rigidity and not by surface tension.
 For the same choice of parameters
in the COP case, instead , the growth process depends on
 two distinct  lengths $L(t)=(2 \Gamma t)^{1/6}$
and $k^{-1}_m(t)$, where $k_m(t)$ is
the wave vector where $C(k,t)$ reaches its maximum and varies in time as
$k_m (t) \sim (\frac{d \ln t}{t})^{1/6}$.
The same mechanism which determines the multiscaling behaviour of the
structure factor upon approaching an ordinary critical
point leads to the
multiscaling form for $\tilde C(k,t) \sim (k_m^{3-d}L^3)^{\phi(k/k_m)}$
with $\phi(x)=(3 x^2 -x^6)/2$.

 For the choice $b=0$, $w=0$ and $T_f >0$
we have investigated numerically the behaviour
of the solutions of eq.(\ref{eq:ct}), with NCOP dynamics.
As shown in Fig. \ref{fig:fig2} by plotting $C(k_m,t)$
times $T_f^{d/(4-d)}$ against time multiplied by $T_f^{4/(4-d)}$
one obtains a remarkable data collapse over several decades.
This suggests the existence of a
typical relaxation time $\tau$ which scales with the temperature $T_f$
according to the law $\tau \sim T_f^{-4/(4-d)}$.
Such a
dynamic scaling relation can be deduced heuristically, by matching
the typical value of the maximum of the structure factor, which
grows as $\sim t^{d/4}$
up to the characteristic time $\tau$,
with the equilibrium value $C_{eq}(k_m)$ at $T_f$
which is proportional to $T_f^{-d/(4-d)}$.

 Let us consider a slightly different case, where
$b=rw/g$, i.e.  a quench to the tricritical point
at ($b \neq 0$, $T_f=0$).
Here the two lengths $k_m^{-1}$ and $\xi$, discussed above, diverge
upon approaching equilibrium.
 One observes a crossover from an initial regime characterized by a
structure factor with a peak at finite values of $k$ to the true asymptotic
regime displaying the ordinary Bragg peak at $k=0$ at late times.
 For the NCOP dynamics one finds from the self-consistency
conditions eqs. (\ref{eq:bt}) and (\ref{eq:qt}) that asymptotically
$2 \Gamma B(t) \sim a t^{1/2}$, with $a>0$, and
$2\Gamma Q(t)=- d \log(t)/4$.
Therefore the structure factor assumes
at late times the scaling form:
$\tilde C(k,t) \sim t^{d/4} \exp(-x^4+2 a x^2)$
where $x=(2 \Gamma t)^{1/4} k$.
 Notice the unusual feature, for NCOP, of
the maximum of the structure factor  at finite wavevector
moving slowly at the rate $\sim t^{-1/4}$ towards the origin .
In other words  initially $B(t)<0$
(for times such that $S_0(t)<|b|/w$),
and the system promotes fluctuations with $k_m(t)>0$, which
are eventually suppressed by the growing of correlations which cause
$k_m \to 0$. However, the two characteristic lengths
$L(t)$ and $k^{-1}_m(t)$ vanish at the same rate and standard scaling
is observed.

A novel behavior is  observed when one crosses the Lifshitz line
and $b_r<0$  since asymptotically
the structure factor $C(k,t)$ displays a peak at finite wavevector, which is
given by  $k_m^2(t)=-B(t)/2t$ .
 The main difference with the cases treated above is the
finite value of $k_m^{-1}$ in the late regime.

For the NCOP dynamics at $T_f=0$ a saddle point estimate
of the integrals yields:
\begin{equation}
\label{eq:s0}
S_0(t)\sim K_d e^{-2 \Gamma Q(t)} e^{\frac{\Gamma B^2(t)}{2 t}}
[\frac{-B(t)}{2t}]^{(d-1)/2}[-B(t)]^{-1/2}
\end{equation}
\begin{equation}
\label{eq:s2}
S_2(t)\sim K_d e^{-2 \Gamma Q(t)} e^{\frac{\Gamma B^2(t)}{2 t}}
[\frac{-B(t)}{2t}]^{(d+1)/2}[-B(t)]^{-1/2}
\end{equation}
with $K_d=[2^{d-1} \pi^{d/2} \Gamma(d/2)]^{-1}$.
Upon requiring that $S_0(t)$ and $S_2(t)$ reach a constant value at equilibrium
we obtain:
\begin{equation}
\label{eq:q0}
Q(t)\sim \frac{ B^2(t)}{4t}-\frac{1}{4 \Gamma} \log[-B(t)]
+\frac{(d-1)}{4 \Gamma} \log[\frac{-B(t)}{2t}]
\end{equation}
 For consistency, since in this region for large times $\dot B(t) <0$ and
$\frac{S_2(t)}{S_0(t)} \to k_m^2$
we find asymptotically $B(t) \sim -2 k_m^2 t$.
After the quench from a high temperature initial state
the structure factor develops a peak at finite values of $k$,
corresponding to a quasi-ordered pattern, with periodicity given
by $\frac {2 \pi} {k_m}$.
The height of the peak grows in time according to the power law $t^{1/2}$.
This is due to the competition between
curvature and surface tension controlled
fluctuations which eventually lead to a layered structure at $T_f=0$.
Within this regime we find the following asymptotic form of the
structure factor
\begin{equation}
\label{eq:peak}
C(k,t) \sim \sqrt{t} \exp{-2 \Gamma (k^2+ \frac{B(t)}{2t})^2 t}
\end{equation}
which obeys the scaling form $C(k,t) \sim t^{\alpha / z} f(|k-k_m| t^{1/z})$
with $\alpha=1$ and $z=2$.
This unusual lack of dependence of the structure factor on the
dimensionality, d,
reflects the fact that the largest fluctuations characterizing the
ordering process have a finite wavenumber and thus the density of state
contribution does not manifest itself.
  We have considered also the finite temperature behavior
of eq.(\ref{eq:ct}) and
found data collapse for the curves representing the height
of the peak times $T_f$ versus the time multiplied by  $T_f^2$.
In fact, the height of the peak
grows according to eq.(\ref{eq:peak}) as $t^{1/2}$,
while  for $T_f>0$ a simple calculation shows that
the structure factor settles at a finite
value $\tilde{C}_{eq}(k_m) \sim T_f^{-1}$.
By matching the two curves one finds a
relaxation time $\tau$  proportional to $T_f^{-2}$.

 In the case of the COP dynamics the only qualitative difference
observed is the weak dependence of the peak position on time,
since $k_m(t)$
approaches as $\log(t)/t$ its equilibrium value. In this
case, therefore, the conservation law turns out to be an irrelevant
constraint since it acts at $k=0$, while the most important fluctuations
are grown at finite $k$.

 A natural question is to ask ourselves how generic is the
$t^{1/2}$ behaviour we have observed, namely whether the present
results for $N \to \infty$ have any relevance with respect to other
models. We believe that our findings are consistent
with the very late time behavior ($z=2$) observed in numerical simulations by
Elder et al. \cite{VI} in a study of the NCOP Swift-Hohenberg
equation with $N=1$ and $d=2$ within the structured
phase.
On the other hand, a part
for a special choice of parameters, we do not find any evidence
in our model of their growth exponent $z=4$ related to curvature
relaxation, a feature probably due to the lack of topological defects.
 We finally remark that the dynamical equations we solved are
very closely related to those derived by employing the
dynamical Hartree approximation for $N=1$ \cite{Daw}.

\begin{figure}
\caption{Phase diagram for $r \le 0$ and $d=3$}
\label{fig:fig1}
\end{figure}
\begin{figure}
\caption{ Data collapse for the peak of the structure
($T_f=0.1,\ 0.05, \ 0.01, \ 0.005, \ 0.001 $)
factor at different temperatures for $d=3$, $b=0$, $r=-0.1$ , $g=1$ , $w=0$}
\label{fig:fig2}
\end{figure}
\end{document}